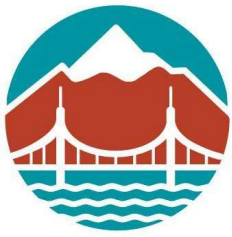

# 2025 EERI LFE Travel Study – Mexico:
## Lessons in soft soils, subsidence, and site effects

M. D. Sanger[1], K. Buyco[2], M. S. Ibrahim[3], P. S. Ramos[4], and A. A. Acosta[5]

**ABSTRACT**

The 1985 M8.1 Mexico City earthquake marked a turning point in Mexican earthquake engineering, underscoring the influence of soft soils, subsidence, and site effects on seismic performance in the Valley of Mexico. In the four decades since, both research and engineering practice have evolved significantly, shaped by subsequent events such as the 2017 M7.1 Puebla-Morelos earthquake. Integrating observations made during the Learning from Earthquakes Travel Study program and desk study findings, this paper summarizes the progression of geotechnical knowledge and practice in Mexico City, through the lessons of 1985 and 2017, through the lens of site effects, soil zonation, subsidence and emerging directions and future trends.

## Introduction

In the four decades since the 1985 M8.1 Mexico City earthquake, the earthquake engineering community's understanding of the regional geotechnical setting and seismic risk has evolved significantly. This event, along with other earthquake events in the region such as the 2017 M7.1 Puebla-Morelos and 2017 M8.2 Oaxaca-Tehuantepec earthquakes, highlighted the critical influence of soft soils, subsidence, and site effects on seismic performance in the Valley of Mexico. These lessons have shaped both research priorities and engineering practice in the region. This study summarizes the evolution of knowledge on site effects and ground deformation in Mexico City, beginning with insights that predate the 1985 earthquake, continuing through the lessons reinforced by the 1985 and 2017 events, and concluding with a perspective on emerging challenges and future directions. The discussion is informed by both a desk study and field observations made during the Learning from Earthquakes Travel Study program, including visits to Puebla and Mexico City. While we do not claim expertise in Mexican earthquakes or geotechnical conditions, this paper reflects what was learned from the program, with emphasis on the perspectives shared by local practitioners and researchers. To provide context, Figure 1 presents a high-level timeline of the evolution of understanding of the region's seismic geotechnical issues which was synthesized from the program. The paper is organized chronologically and focuses on Mexico City for brevity, tracing developments in geotechnical earthquake resilience through site effect characterization (i.e., "zonation") and subsidence (e.g., differential settlement) over the past forty years.

---

[1] Graduate Student, Civil and Environmental Engineering, University of Washington, Seattle, WA 98195 (email: sangermd@uw.edu)
[2] Senior Risk Modeler, SageSure, San Francisco, CA 94534
[3] Graduate Student, Civil and Environmental Engineering, McGill University, Montreal, QC H3A 0C3
[4] Graduate Student, Civil and Environmental Engineering, University of Aveiro, Aveiro 3810-104
[5] Design Engineer, Degenkolb Engineers, San Francisco, CA 94103

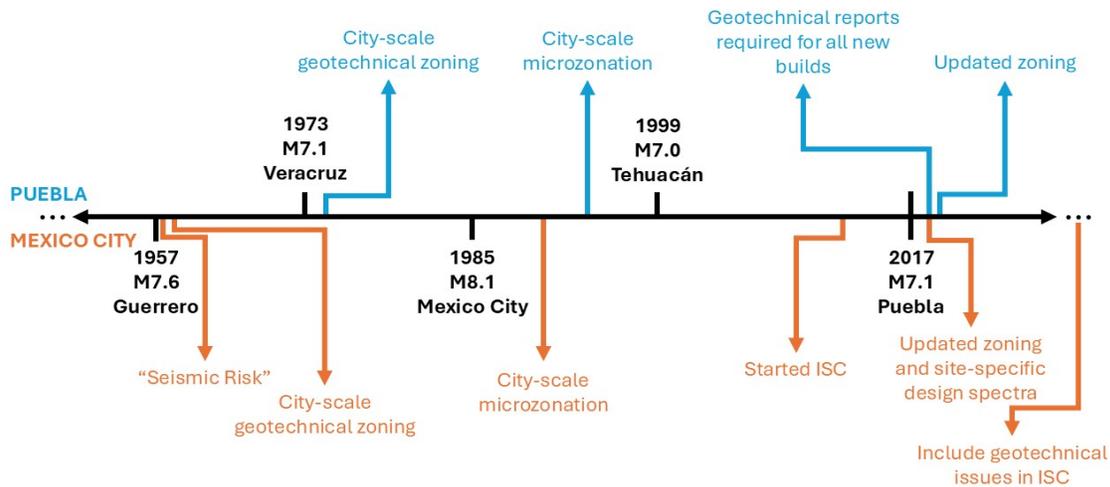

Figure 1. Timeline of large-magnitude earthquakes in central Mexico and the geotechnical practice and policy changes that they precipitated.

**Before 1985 Mexico City earthquake**

The unique soil conditions in Mexico City have important implications for earthquake engineering. Most of the city center is built on the sediments of the ancient Lake Texcoco. The area of this ancient lake was 700 km$^2$ during Aztec times but was drained to 140 km$^2$ by the end of the 18$^{th}$ century, and is today almost completely drained [1]. The saturated clay of the lakebed, upon which much of Mexico City is built, is characterized by very low shear wave velocity, $V_s$ (i.e., $V_s$ = 40-90 m/s [1]) and depths of up to 60 m [2]. Consolidation and groundwater extraction have led to dramatic subsidence within the Mexico City lakebed, with settlements of up to 10 m in some areas of the city [2]. The distinctive soil conditions in Mexico City have important implications for site effects and thereby structures in the metropolitan area, as evidenced by the spatial concentration of structural damage in the 1985 Mexico City and 2017 Puebla-Morelos earthquakes [3, 4]. However, prior to the 1985 earthquake, Mexico City was aware of the influence of its unique geological setting on earthquake risk. The 1957 M7.6 Guerrero earthquake caused damage to several buildings located on soft soils in the former lakebed and thus began the journey of modern earthquake engineering in Mexico.

**Site effects and zonation**

Following the 1957 earthquake, the Mexico City Building Code emergency edition proposed three subdivisions within Mexico City: Hill Zone, Transition Zone, and Lake Zone. Seismic design coefficients were highest in the Lake Zone and lowest in the Hill Zone. The zonation map was modified in a separate section of the 1976 code but retained its classification: Hill Zone (I), Transition Zone (II), and Lake Zone (III) [4]. From that point onwards, although the geotechnical divisions have been similar, the current seismic code provides a detailed map of site effects and is included to compute the seismic demand.

**Subsidence**

Mexico City's land subsidence was first recognized by Roberto Gayol (1925), who compared precision leveling of 1877 and 1924 near the Cathedral and attributed the phenomenon to the newly built drainage system [5]. Cuevas (1936) later reported in the geotechnical literature that subsidence could be related to groundwater extraction [6]. Scientific confirmation came with Carrillo (1948), who applied Terzaghi's consolidation theory and showed that depressurization of the lacustrine aquitard due to groundwater withdrawals was the primary cause [7]. Pumping rates rose sharply during the 20th century, from 0.5 m³/s in 1910 to 1.5 m³/s in 1930, 6 m³/s in 1940, 9 m³/s through the 1950s–1970s, and 12 m³/s by 1974 [8]. This excessive withdrawal drove potentiometric declines and compaction of the aquitard. The city center exhibited three characteristic phases before 1985: (i) subsidence of 8 cm/yr between 1935 and 1948, (ii) accelerated subsidence of 29 cm/yr between

1947 and 1958 during aquitard dewatering, and (iii) moderated subsidence of 5-9 cm/yr after 1959 following the capping of central wells that directly drained the upper aquitard [9]. By 1985, cumulative subsidence in the historic core reached 7.5 m, with an associated 32 m potentiometric decline [9].

## 1985 Mexico City earthquake

The 19 September 1985 M8.1 Mexico City earthquake occurred along the subduction interface off the Pacific coast of Mexico, approximately 350 km southwest of Mexico City. While shaking was severe in the coastal states of Mexico and Guerrero, the information of impacts were only available in Mexico City. The disaster underscored the influence of site conditions on earthquake impacts and motivated revisions to zonation.

**Site effects and zonation**
In 1985, the damaged buildings were concentrated in the northwest area of Mexico City in the Lake Zone. The damaged buildings were mostly founded in soil with predominant periods from 2 to 3 s (basin depths 20-35 meters), and rock motions (e.g., CUP5 station) recorded from this earthquake showed spectral acceleration peaks in between 1 to 3 s. Buildings with effective periods of approximately 2 s were especially vulnerable to severe damage, with amplifications of up to 5 times were observed [1]. As a result, it was acknowledged that the seismic design criteria in the Lake and Transition Zones needed to be strengthened. The 1985 emergency code was established five weeks later and significantly increased the seismic design coefficients in Zone II and Zone III. The 1987 code further updated these seismic design coefficients [4]. The 2004 code further subdivided Zone III (i.e., IIIa, b, c, d) to reflect the variation of basin depths and soil conditions. The current code in Mexico City has a detailed site effect micro-zonation and it is used to quantify the seismic demand.

**Subsidence**
In 1989, after the accumulated damage from the 1985 earthquake, long-term differential settlement, and regional subsidence, the Metropolitan Cathedral in Mexico City was put into a formal salvage program [10]. Differential settlement started during construction because part of the site sat on ground pre-consolidated by Aztec structures while adjacent areas were softer lacustrine clays [10]. By the late 19$^{th}$ century, the differential between the western tower base and the apse was 1.5 m; leveling records since 1907 show rates rising from a few cm/yr to more than 20 cm/yr, peaking above 30 cm/yr in the 1940s [10]. Rates dropped after wells were banned in the central part of the city in the early 1960s, then rose again in the early 1980s [10]. By October 1989 the differential reached 2.4 m, triggering the well-known underexcavation program [10].

## 2017 Puebla-Morelos earthquake

The 19 September 2017 M7.1 Puebla-Morelos earthquake occurred within the subducted Cocos Plate at a depth of approximately 50 km, with an epicenter in the state of Morelos, approximately 120 km southeast of Mexico City. Despite its smaller magnitude relative to 1985, the significant losses in densely populated areas again reaffirmed the importance of characterizing site effects and their evolution through subsidence.

**Site effects and zonation**
In 2017, collapsed buildings were concentrated in the western basin margin (i.e., Zones 3a and 3b), where site effects and input ground motion characteristics amplified damage. These locations have predominant soil periods of 1 to 2 s (basin depths 10-35 m), overlapping with rock motion peaks at 0.5 to 1.5 s [1]. As in 1985, resonance between soil and input motions explains the concentration of collapses, though the shorter periods in 2017 disproportionately affected short- and mid-rise buildings [4]. Basin-edge effects, which generate and propagate surface waves, may have further amplified motions in these areas [1, 4].

**Subsidence**
Groundwater-driven subsidence in Mexico City has caused a measurable shortening of site periods, especially where clays are thick [11]. Post-2017 HVSR (horizontal-to-vertical spectral ratio) surveys confirmed the trend is accelerating, with subsidence driving a continual shift in basin resonance [12]. As site periods shorten, amplification peaks shift, increasing seismic demand on some buildings while reducing it for others. This evolving alignment has been linked to heightened vulnerability and, in some cases, collapse [13]. During the 2017 earthquake, pre-existing subsidence-related settlement and tilt, evolving site response (i.e., period shifts), and soil-structure interaction were major contributors to collapse. A post-event synthesis attributes about two-thirds (63%) of the 38 Mexico City collapses to these geotechnical drivers [12]. At the Plaza de las Tres Culturas (Tlatelolco), cracks opened in the nave vaults and roof of the Church of Santiago (Figure 2b,c). During the visit, the priest reported that one tower is out of plumb and has experienced differential settlement (Figure 2b,d). The church was temporarily closed for evaluation and repairs. Similar behavior is visible in the adjacent residential blocks, where differential settlement is apparent (Figure 2a).

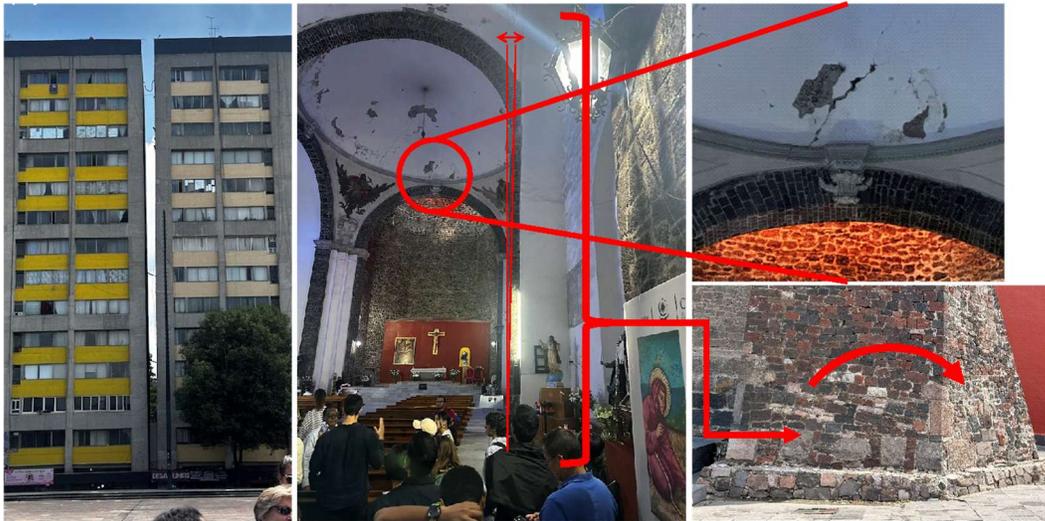

Figure 2. Observations of differential settlement in Tlatelolco, including (a) adjacent residential buildings showing relative lean; (b) interior of the Church of Santiago with vault cracking at the nave and out-of-plumb column; (c) cracks at the cornice; and (d) rotational settlement of the tower.

## Moving towards the Future

Open-source information is increasingly important for advancing geotechnical engineering, enabling reproducibility, and improved hazard models [14]. Countries such as the United States [15] and the United Kingdom [16], among many others, are in the early stages of open geotechnical data repositories. Mexico has made significant advances in open-access geological data (e.g., INEGI, Servicio Geológico Mexicano) and are thus well positioned to build on its current efforts by expanding open-source geotechnical platforms for the future of site effects, subsidence monitoring, and earthquake hazard analysis.

## Conclusions

Over the past forty years, Mexico City has significantly advanced its understanding of site effects and subsidence, leading to important changes in building codes, seismic zonation, and engineering practice. Despite these improvements, groundwater-driven subsidence continues to alter site periods and reshape seismic demand, creating evolving risks for both new and existing structures. While challenges remain, Mexico City today is better prepared and more resilient than it was in 1985, with continued progress depending on integrating science, practice, and data sharing to meet future seismic hazards.

## Acknowledgments

This study was conducted as a part of the Learning from Earthquake's Travel Study Program to Mexico in September 2025, which is a project of the Earthquake Engineering Research Institute (EERI). The program was made possible by support from the EERI Learning from Earthquakes Endowment Fund, the Sociedad Mexicana de Ingeniería Sísmica (SMIS), and the 2025 trip's Anchor Sponsor, Computers and Structures, Inc.